\documentclass{article}
\usepackage{spconf,amsmath,graphicx}
\usepackage{multirow}
\usepackage{booktabs}
\usepackage{makecell, tabularx}
\usepackage{url}
\usepackage{paralist}

\usepackage{arydshln}

\title{Investigating neural audio codecs for speech language model-based speech generation}
\makeatletter
\def\@name{\textit{Jiaqi Li$^{2,\star,\dag}$ \thanks{$\star$ The first two authors contributed equally to this work.} \thanks{\dag \hspace{0.2pt} Work done during internship at Microsoft.}, Dongmei Wang$^{1,\star}$, Xiaofei Wang$^1$, Yao Qian$^1$, Long Zhou$^1$, Shujie Liu$^1$}\\ 
\textit{Midia Yousefi$^1$, Canrun Li$^1$ Chung-Hsien Tsai$^1$, Zhen Xiao$^1$, Yanqing Liu$^1$}\\
\textit{Junkun Chen$^1$, Sheng Zhao$^1$, Jinyu Li$^1$, Zhizheng Wu$^2$, Michael Zeng$^1$}\\
}
\makeatother
\address{$^1$ Microsoft, One Microsoft Way, Redmond, WA, USA \\
$^2$ The Chinese University of Hong Kong, Shenzhen, China}

\begin{document}
%
\maketitle
\begin{abstract}
Neural audio codec tokens serve as the fundamental building blocks for speech language model (SLM)-based speech generation. However, there is no systematic understanding on how the codec system affects the speech generation performance of the SLM. In this work, we examine codec tokens within SLM framework for speech generation to provide insights for effective codec design. We retrain existing high-performing neural codec models on the same data set and loss functions to compare their performance in a uniform setting. We integrate codec tokens into two SLM systems: masked-based parallel speech generation system and an auto-regressive (AR) plus non-auto-regressive (NAR) model-based system. Our findings indicate that better speech reconstruction in codec systems does not guarantee improved speech generation in SLM. A high-quality codec decoder is crucial for natural speech production in SLM, while speech intelligibility depends more on quantization mechanism.

\end{abstract}
\begin{keywords}
neural audio codec, speech language model, speech generation, tokens, codec investigation
\end{keywords}
\section{Introduction}
\label{sec:intro}

The emergence and success of large language models, such as the GPT series of work \cite{GPT2020Brown, GPT4}, have inspired research in the field of speech language model (SLM) within the speech generation community \cite{audioLM2023, valle2023, soundstorm2023, audioPalm2023}. Rather than synthesizing the speech in a sample-by-sample fashion \cite{wavenet2016} or by estimating the continuous features such as mel-spectrum, SLM directly predict the discrete speech tokens. These predicted tokens are then used by the pre-trained decoder module to reconstruct the waveform. By modeling discrete speech tokens within the language model framework, we can harness the advancements from large language models to enhance speech generation tasks. Another advantage of using discrete tokens is that it facilitates the construction of multi-modal model. These tokens originating from various modalities (such as speech, text, video, etc.), can be seamlessly integrated and processed in a unified manner. 

Discrete speech representation tokens primarily stem from two lines of work: self-supervised learning \cite{vq-wav2vec, hubert, w2v-bert}, resulting in semantic tokens, and the neural audio codec system \cite{soundstream, encodec, dac}, yielding codec tokens. Initially, semantic tokens were adopted by SLM-based speech generation \cite{gslm_2021}. While they produce intelligible speech, they fail to capture consistence acoustic traits like speaker identity. In contrast, the neural audio codec system, which is based on residual vector quantization (RVQ), generates hierarchical codec tokens that preserve rich acoustic information. These codec tokens are then employed, either alone \cite{ valle2023} or alongside semantic tokens by SLM to generate speech \cite{audioLM2023, soundstorm2023}. 

Neural audio codec systems were originally designed for communication purposes, whereas compression rate and reconstruction quality are the primary evaluation metrics. Extended from VQ-VAE \cite{vq-vae} and introducing the RVQ module, SoundStream \cite{soundstream} stands as the pioneering neural audio codec model. It comprises three essential modules: encoder, RVQ-based quantizer, and decoder. These components are jointly trained using both reconstruction and adversarial losses. Subsequently, a series of neural audio codec models are proposed \cite{encodec, dac, vocos}, expanding upon SoundStream. 
Encodec \cite{encodec} incorporated LSTM layers and a small transformer language model over the quantized units to further reduce bandwidth. 
Vocos \cite{vocos} predicted the short-time Fourier transform (STFT) coefficients instead of waveform in the decoder stage, which demonstrates improvement over Encodec. DAC \cite{dac} proposed two key improvements over Encodec. First, it addressed the codebook collapse problem by reducing the dimension of the latent vector to a small value for quantization. Second, DAC replaced the ReLU activation function with the snake activation function \cite{snake-actiFunc}, offering benefits for reconstructing periodic signals such as speech and music. Lately, there has been a surge in research efforts focused on designing low bit-rate codec system \cite{NaturalSpeech3_2024, semantiCodec, promptCodec2024, timeInvCodec2024, HILCodec2024, singleCodec2024} that can be integrated with speech language models, aiming for greater efficiency. A comprehensive review of existing neural codec models and the audio language models can be found in \cite{CodecSuperb2024, reviewALM2024}. 

While the original goal of neural codec design is to achieve a high compression rate and superior signal level reconstruction quality, it remains unclear how the codec affects the SLM-based speech generation.
In this work, we investigate the effectiveness of the established codec systems within the framework of SLM for zero-shot speech generation. Our goal is to identify key components for designing codec systems tailored to SLM. Regarding codec systems, we select three high-performing models: Encodec \cite{encodec}, Vocos \cite{vocos} and DAC \cite{dac}. To ensure a fair comparison, we retrain these models and their variants by using the same data set and loss functions. We consider two SLM-based zero-shot speech generation models. First, we adopt the masked-based parallel speech generation model, originally proposed in the SoundStorm work \cite{soundstorm2023}. We built upon the original design by incorporating enhancements such as mask span and classifier-free guidance (CFG) techniques \cite{maskedAudioGen2024, CFG2022}. Second, we explore the AR + NAR models introduced in VALL-E work \cite{valle2023}. We systematically assess both the speech reconstruction quality of the codec systems and the speech generation quality of the speech generation systems. The main contributions of this work include:

\begin{compactitem}
    \item We leverage large-scale speech data to retrain high-performing codec models for a unified comparison. 
    \item We integrate these codecs into two popular SLM-based speech generation systems for investigation. 
    \item Thorough evaluation and analysis inform effective codec design for SLM-based speech generation. 
\end{compactitem}

\section{Neural audio codec systems}
\label{sec:neuralCodec}

\begin{figure}[t]
    \centering
    \includegraphics[scale=0.21]{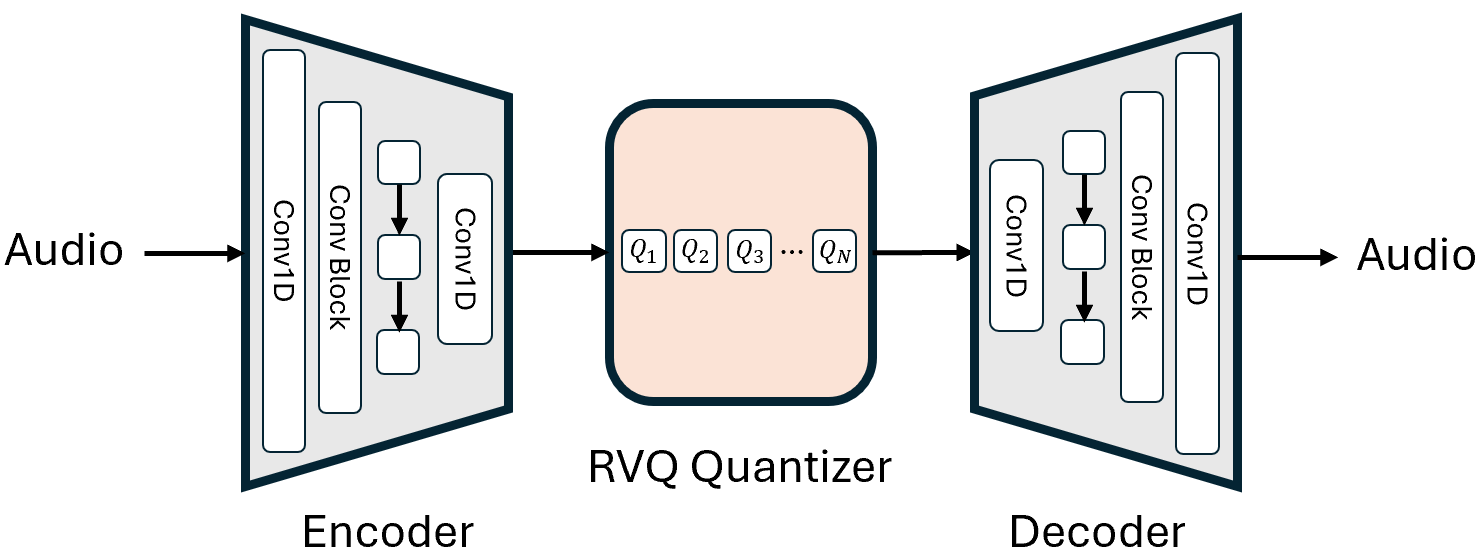}
    \caption{High-level architecture of neural audio codecs}
    \label{fig:codec}
\end{figure}

Neural audio codecs can compress audio into discrete representations used by speech generation models. 
Figure \ref{fig:codec} presents a high-level architecture of a neural audio codec. It has an encoder that downsamples waveforms to a much lower sampling rate (e.g., 50Hz), a residual vector quantization (RVQ) module that discretizes latent features, and a decoder that reconstructs audios from discrete tokens. In the following sections, we examine the design choices of Encodec \cite{encodec}, Vocos \cite{vocos} and DAC \cite{dac}.
We categorize two aspects of a codec that affect speech generation: i) \textbf{vector quantization (VQ) scheme} which can affect the distribution of tokens and indirectly affect speech modeling complexity; ii) \textbf{decoder scheme} which affects the generated audio quality.
\vspace{-1mm}
\subsection{Encodec}
\vspace{-1mm}
\label{sec:encodec}
For quantization, Encodec \cite{encodec} uses Residual Vector Quantization (RVQ) similar to SoundStream~\cite{soundstream}. 
The output of the $n$-th quantizer $Q_n$ is expressed as 
    $Q_n = \text{VQ} (Q_{0} - \sum_{i=1}^{n-1}Q_i)$
where $Q_0$ is the continuous latent vector. 
The VQ operation at each layer is to find the codebook vector that is closest to the residual embedding in Euclidean distance. 
During training, the codebook vectors are updated using Exponential Moving Average (EMA)\cite{vq-vae}. To mitigate the codebook collapse problem, Encodec also applies a ``restart" technique to replace unused codebook vectors with candidates sampled from the batch. 
Encodec utilizes a fully-convolutional decoder SEANet\cite{seanet} same as SoundStream\cite{soundstream}, with transposed convolutions to upsample the quantized features into waveform.
Two small LSTM layers are added to improve sequence modeling.

\subsection{Vocos}
\vspace{-1mm}
\label{sec:vocos}
Vocos \cite{vocos} is a GAN-based Vocoder trained to produce STFT coefficients. It can be integrated into any neural codec framework as a decoder \cite{vocos}, either by training from a frozen pre-trained encoder and quantizer, or by building it from scratch in an end-to-end manner.
Vocos predicts STFT coefficients (logarithmic scale spectrum amplitude and the phase values) instead of raw waveforms, and upsampling to waveform is realized through inverse Fourier Transform \cite{vocos}.
This system has shown to produce higher quality audios than original Encodec \cite{vocos}, and is used in speech generation system VALL-E 2 \cite{valle2}.


\subsection{DAC}
\vspace{-1mm}
\label{sec:dac}
Descript-audio-codec (DAC) \cite{dac} is a recent codec system that features several VQ improvements over Encodec.
DAC mitigates codebook collapse by quantizing in a very low-dimensional latent space. Its updated RVQ is: $
    Q_n = \text{Proj\_Out}(\text{VQ} (\text{Proj\_In}(Q_{0}) - \sum_{i=1}^{n-1}\text{Proj\_In}(Q_i)))$
where $\text{Proj\_In}$ is a linear 
projection from the original latent space (1024 dimensional) to the low-dimensional quantization latent space (8 or 32 dimensional). 
It also changes the VQ lookup distance from Euclidean distance to cosine similarity for stability. 
DAC uses an explicit MSE codebook loss function instead of the EMA update scheme to learn the projection functions.
This loss can be expressed as $L_{codebook} = ||q - z||_2^2$, where $z$ and $q$ denote the looked-up vectors and the quantized vectors, respectively.

Compared with the decoder in Encodec, DAC replaces the ReLU activation with Snake activation, which is shown to benefit periodic signal reconstruction quality \cite{snake-actiFunc}.

\section{Speech generation with codec tokens}
\label{sec:speechGen}


For the speech generation task, we selected two types of SLM-based systems. The first one is a masked-based parallel speech generation model \cite{soundstorm2023} conditioned on oracle semantic tokens. The second is an AR + NAR models-based text-to-speech system \cite{valle2023}. Both systems predict the codec tokens for the target speech which are then used to reconstruct the final waveform via a pre-trained codec decoder.

\begin{figure}[t]
    \centering
    \includegraphics[scale=0.55]{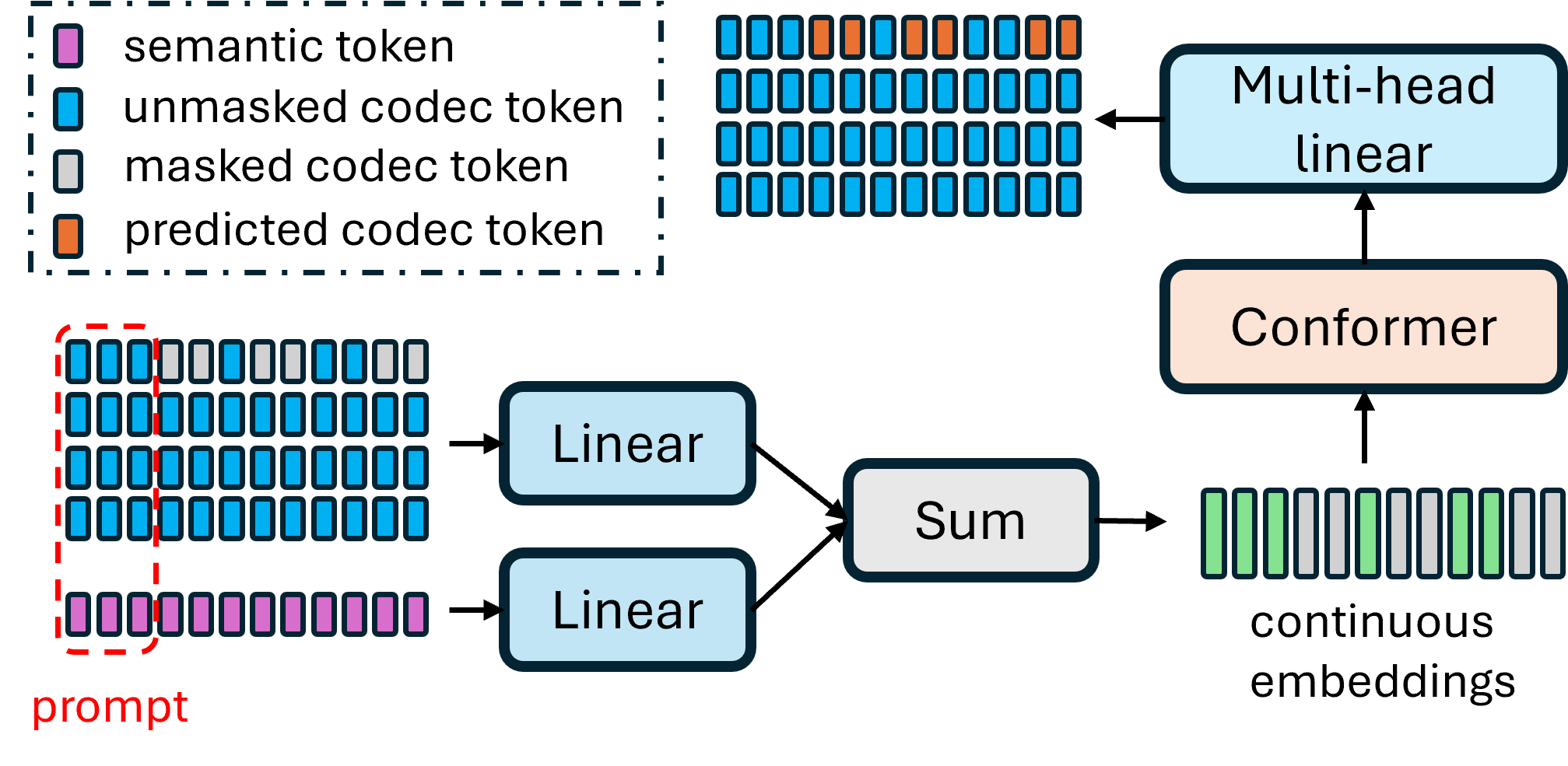}
    \vspace{-8mm}
    \caption{Model architecture of masked-based parallel speech generation.} 
    \label{fig: maskParaGen} 
\end{figure}

\subsection{Masked-based parallel speech generation}
\label{subsec:maskParalledlGen}

The masked-based parallel speech generation is proposed in \cite{soundstorm2023} which is inspired by maskGIT in image generation field \cite{maskGIT2022}. Unlike the AR-based method, masked-based parallel speech generation produces codec tokens for the entire speech sample in a batch-style manner, iterating through multiple rounds based on confidence scores. 
Fig.~\ref{fig: maskParaGen} illustrates the model architecture for masked-based parallel speech generation. The backbone employs a bidirectional self-attention-based Conformer, which predicts masked codec tokens using summed embeddings of codec tokens and semantic tokens. Due to the hierarchical structure of the RVQ-based codec tokens, masked-based parallel generation occurs layer by layer, advancing to the next layer only when all tokens from the current layer have been estimated.   

We adopted the span-based masking strategy \cite{maskedAudioGen2024} where a sequence of block-wise (block size is set as 5 in our case) masks is applied instead of individually masking each token in every iteration. To further enhance the quality of the generated speech, we integrated the annealing-based CFG mechanism, as proposed in the same work \cite{maskedAudioGen2024}. During training, the model is trained both conditionally and unconditionally with a certain probability. During inference, the generated signal is sampled from a linear combination of the predicted conditional and unconditional probabilities with ratio controlled by the masking rate.    
This mechanism gradually steers the generation process from being solely guided by semantic tokens to incorporating contextual infilling. Ultimately, the predicted codec tokens are transformed into speech waveforms using the decoder module of the codec system.  

Note that in this experiment, our focus is solely on investigating the framework of masked-based parallel generation across various codec systems. We achieve this by utilizing oracle semantic tokens as input, without the need for text-to-semantic-tokens mapping as in \cite{soundstorm2023}.

\subsection{Speech generation with AR and NAR models}
\label{subsec:ARNAR_speechGen}

\begin{figure}[t]
    \centering
    \includegraphics[scale=0.55]{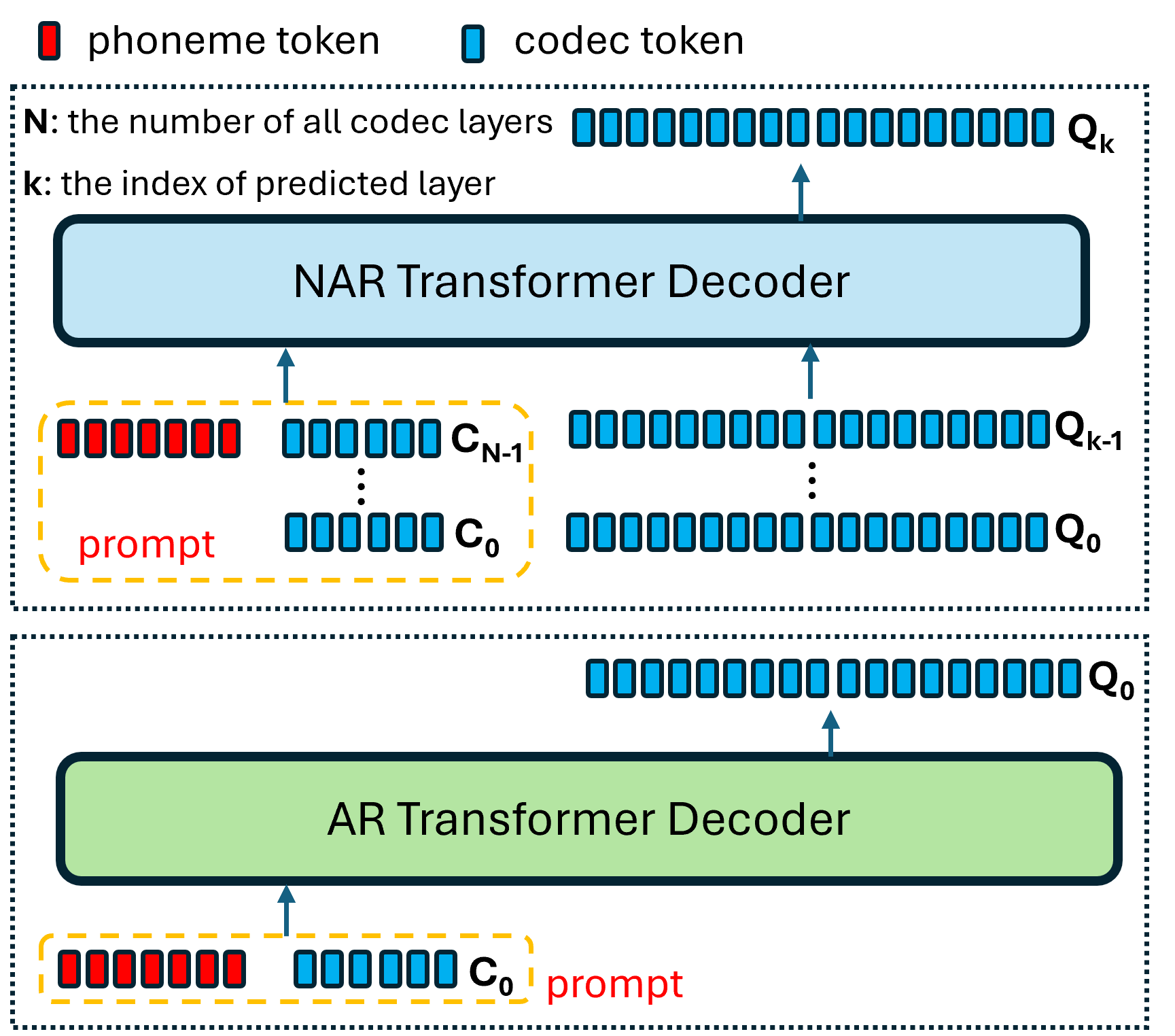}
    \vspace{-3mm}
    \caption{AR + NAR speech generation.} 
    \label{fig: valle} 
     \vspace{-2mm}
\end{figure}

The alternative SLM-based speech generation system we've experimented with employs AR and NAR models, as proposed in the VALL-E work \cite{valle2023}. Fig.~\ref{fig: valle} illustrates the model architectures of the method. The overall system involves two stages of generation. First, the AR model takes the phoneme sequences derived from the text and the prompt codec tokens from the first quantization layer as input, predicting the first layer of the codec tokens for the target speech in an AR manner. Subsequently, the NAR model predicts the remaining codec tokens layer-by-layer, based on all the already predicted layers of codec tokens combining with the phoneme sequence and the prompt codec tokens of all quantization layers, in a parallel fashion. Both the AR and NAR models utilize the same Transformer model architecture. However, the AR model operates causally, whereas the NAR model operates in parallel.

\section{Experiment}
\label{sec:exp}

\subsection{Experiment of codec reconstruction}
\label{subsec:exp-codec}

%
\subsubsection{Model configuration}
\label{subsubsec:modelCFG-codec}

We evaluated official 24kHz pretrained models of Encodec~\cite{encodec}, Vocos (with Encodec features)\cite{vocos}, and DAC~\cite{dac}. 
We also evaluated these reproduced codec models. Since the official Encodec does not provide training codes, we reproduced a baseline codec with encoder and decoder architecture from Encodec. For the quantizer, we used an EMA quantization module
with no restart technique. After obtaining the baseline codec system, we trained a Vocos decoder with the encoder and quantizer frozen.
We reproduced DAC using its official repository\footnote{https://github.com/descriptinc/descript-audio-codec}. Then, we trained a Vocos decoder for the reproduced DAC with frozen encoder and quantizer.

We retrained all the codec models with 54k hours of the Librilight-Large dataset \cite{librilight}. 
Table \ref{tab:codec-training-comparison} shows a comparison between official pre-trained models and our reproduced models.
We trained our models using 16kHz sampling rates to align with our training set.
We also used a fixed bitrate for training reproduced models. 

\begin{table}[t]  
    \centering
    \caption{Training configuration comparison between official pre-trained models and our reproduced models.}
    \resizebox{\linewidth}{!}{
    \begin{tabular}{ccccc}
    \toprule 
        \multirow{2}{*}{Model name} & Quantization  & Sampling & Token & Bitrate \\
        &method& rate (kHz)&rate (Hz) & (kbps)\\   \hline
         Encodec-official & EMA w/ restart & 24 & 75 & 1.5-24 \\
        Vocos-official & - & 24 & 75 & 1.5-12 \\ 
        DAC-official & Projection & 24 & 75 & 0.75-24 \\ 
       \hdashline[1pt/2pt]\hdashline[0pt/1pt] 
        Baseline-16kHz & EMA & 16 & 50 & 4 \\ 
        Baseline-Vocos-16kHz & - & 16 & 50 & 4 \\ 
        DAC-16kHz & Projection & 16 & 50 & 4 \\
        DAC-Vocos-16kHz & - & 16 & 50 & 4 \\ 
        \bottomrule
    \end{tabular}
    }
    \label{tab:codec-training-comparison}
\end{table}






\subsubsection{Training Details}
\label{subsubsec:training-details-codec}

For discriminators, we used a combination of a multi-scale STFT (MS-STFT) discriminator from Encodec \cite{encodec}, and a multi-period discriminator (MPD) from HiFi-GAN \cite{hifigan}. 
We used the same discriminator implementation for all reproduced codecs.
For the loss formulation, we followed an effective configuration in \cite{dac} and used a combination of reconstruction loss, adversarial loss and commitment loss.

Each reproduced model was trained on 8 V100 GPU for 200k steps. We used a segment length of 1 second, and a batch size of 22 per GPU. We used AdamW optimizer with learning rate 1e-4, $\beta_1 = 0.8$, $\beta_2 = 0.9$, and an exponential learning rate decay with $\gamma = 0.999996$.

\subsubsection{Evaluation Metrics}
\label{subsubsec:evaluation-codec}
We used the short-split of Librispeech-test-clean \cite{librispeech} as the test set with a duration range spanning from 4 seconds to 10 seconds.
To evaluate the speech quality of the codecs, we used the Perceptual Evaluation of Speech Quality (PESQ) \cite{pesq}, Short Term Objective Intelligibility (STOI) \cite{stoi}, Virtual Speech Quality Objective Listener(ViSQOL) \cite{visqol}, Mel Cepstral Distortion (MCD) \cite{mcd} as indicators.
We also measured the speaker similarity (SIM) between the original speech and reconstructed speech using the WavLM-TDNN model \cite{Chen2021WavLM}.
For Word Error Rate (WER) evaluation, we used a market-leading ASR API to get the transcripts and calculate word error rates with ground-truth transcripts. 

\subsubsection{Results}
\label{subsubsec:results-codec}
As shown in Table \ref{tab:results-codec}, for official models, we observed that the official DAC model performs the best among most metrics. 
We also observed that using the Vocos-official decoder for Encodec improves its codec quality.
Our reproduced models share similar trends, where the reproduced DAC-16kHz model performed better than the baseline codec. Also, adding Vocos can improve some aspects of sound quality, in particular, we found this gives better speaker similarity and ViSQOL scores, and lower word error rates. 
We observed that our reproduced models generally outperform official models in the evaluation metrics. This superiority may stem from both our training set and test set being within the audiobook domain, ensuring better alignment between train-test distributions compared to official models, which incorporate crowdsourced data like Common Voice \cite{common-voice} in training. 


\begin{table}
\centering
\caption{Evaluation results of both official codec and retrained codec models}
\vspace{.3em}
\label{tab:results-codec}
\resizebox{\linewidth}{!}{
\begin{tabular}{ccccccc}
\toprule 
Codec models    & PESQ          & STOI          & VISQOL        & MCD($\downarrow$)  & SIM           & WER ($\%$)           \\ \midrule 
Encodec-official & 3.12          & 0.94          & 4.37          & 2.60 & 0.89          & 1.31          \\
Vocos-official   & 3.57          & 0.95          & 4.41          & 2.50 & 0.90          & 1.31          \\
DAC-official     & 3.77          & 0.95          & 4.36          & 2.34 & 0.90          & 1.27          \\  \hdashline[1pt/2pt]\hdashline[0pt/1pt] 
Baseline-16kHz    & 3.63          & 0.95          & 4.44          & 2.32 & 0.90          & 1.29          \\
Baseline-Vocos-16kHz      & 3.62          & 0.95          & 4.47          & 2.58 & 0.92          & 1.20          \\
DAC-16kHz        & \textbf{3.99} & \textbf{0.97} & 4.53 & \textbf{1.95} & 0.94          & 1.12          \\
DAC-Vocos-16kHz  & 3.98          & \textbf{0.97} &             \textbf{4.54}  & 2.00 & \textbf{0.95} & \textbf{1.06} \\ \bottomrule 
\end{tabular}
}
\end{table}

\subsection{Experiment of masked-based parallel generation}
\label{subsec:exp-soundstorm}

\begin{table*}[t]
\vspace{-2mm}
\centering
\caption{Results of masked-based parallel speech generation}
\vspace{.3em}
\label{tab:results-soundstorm}
\resizebox{\linewidth}{!}{
\begin{tabular}{cccccccccccc}
\toprule 
\multirow{2}{*}{ID} & \multirow{2}{*}{Codecs} & \multirow{2}{*}{Token rate} & \multicolumn{3}{c}{Continuation generation}   & \multicolumn{6}{c}{Cross-speaker generation}                                  \\ \cmidrule{4-6} \cmidrule{8-12}
                    &                         &                             & SIM-O         & NISQA         & WER (\%)      && SIM-O         & NISQA         & WER (\%)      & CMOS          & SMOS          \\   \midrule 
GT                  & GroundTruth             & -                           & 0.67          & 3.87          & 0.96          && 0.70          & 3.87          & 0.96          & 0.29          & 4.75          \\  \hdashline[1pt/2pt]\hdashline[0pt/1pt]  
S1                  & Encodec-official        & 75                          & 0.50          & 3.17          & 1.54          && 0.55          & 3.31          & 1.79          & -0.79         & 4.29          \\
S2                  & Baseline-16kHz           & 50                          & 0.59          & 3.33          & 1.27          && 0.58          & 3.52          & 1.85          & -0.38         & \textbf{4.69} \\
S3                  & Baseline-Vocos-16kHz             & 50                          & \textbf{0.61} & 3.47          & \textbf{1.22} && 0.58          & 3.76          & 1.86          & -0.31         & 4.52          \\
S4                  & DAC-16kHz               & 50                          & 0.54          & \textbf{3.80} & 1.37          && 0.58          & \textbf{3.99} & \textbf{1.78} & -0.09         & 4.43          \\
S5                  & DAC-Vocos-16kHz         & 50                          & 0.54          & 3.74          & 1.28          && \textbf{0.59} & 3.97          & 1.79          & \textbf{0.00} & 4.57          \\   \bottomrule 
\end{tabular}
}
\vspace{-2mm}
\end{table*}

\subsubsection{Dataset}
\label{subsubsec:data-soundstorm}

We trained the masked-based parallel speech generation model with 54k hours of Librilight-Large dataset \cite{librilight}. The data was chunked into a maximum 30s for each training sample.
For inference, we utilized the short-split of LibriSpeech-test-clean \cite{librispeech} data set.

\subsubsection{Model configuration}
\label{subsubsec:modelCFG-soundstorm}

For semantic token extraction, we used the Hubert-base model \cite{hubert}. The semantic and retrained codec tokens were generated at 50 tokens per second, while the official Encodec tokens were at a rate of 75 tokens per second. To align the timing between official Encodec tokens and semantic tokens, we up-sampled the embeddings of the semantic tokens to match the official Encodec token rate. The model has 12 layers of Conformer, each layer has 16 attention heads, 1024 dimension of embeddings, 4096 feedforward dimensions, with a convolution kernel size of 5 and rotary positional embeddings, similar as \cite{soundstorm2023}.  
During decoding, we applied 5 iterations for the first codec layer. For annealing CFG configuration, we set the initial and final guidance coefficients to 0 and 2, respectively. For the rest of the RVQ layers, we used the greedily decoding without iterations. Overall, it requires 12 forward passes to predict the tokens for all 8 RVQ layers. 

The model was trained using Adam optimization, with a batch size of 32. We employ the linear decay learning rate scheduler with 10k steps of warmup and a peak learning rate of 1e-4. The model trained with official Encodec tokens used 10 epochs, while other codec tokens utilized 5 epochs.

\subsubsection{Results}
\label{subsubsec:results-soundstorm}

Table~\ref{tab:results-soundstorm} presents the evaluation results. In continuation generation, we used the first 3 seconds of speech from the same utterance as the prompt. For cross-speaker generation, we employed 3 seconds of speech from a different utterance as the prompt. Our evaluation metrics for continuation generation include SIM-O \cite{voicebox2023}, NISQA score \cite{nisqa2021} and WER. Additionally, for cross-speaker generation, we included CMOS (comparative mean option score) and SMOS (similarity mean option score) to assess human perception of speech naturalness and speaker similarity in the generated speech. The testing procedure of CMOS and SMOS followed the same protocol as described in \cite{valle2023}, with 10 subjects participating each test and 10 randomly selected samples for each condition.  

In general, generated speech using the retrained codec tokens outperformed the official Encodec tokens across most evaluation metrics. The WER results of generated speech were similar across all codec tokens, likely because we used oracle semantic tokens as conditions. From NISQA scores, we observed the following trends: i) The Vocos decoding outperformed the waveform-based decoding in the baseline codec model; ii) DAC decoding exhibited similar performance to Vocos decoding, with both surpassing the baseline decoding. Moreover, the CMOS score in cross-speaker generation showed a robust correlation with the NISQA score. These results suggested that employing the snake activation function for waveform decoding had a similar impact on speech naturalness as predicting the STFT coefficients. Comparing Baseline-Vocos-16kHz and DAC-Vocos-16kHz, we observed that the quantization of DAC yields better speech naturalness. 

In terms of SIM-O results, there was a slight discrepancy between continuation generation and cross-speaker generation when comparing the baseline-16kHz and DAC-16kHz. However, considering that all SIM-O scores fell within a narrow range, the discrepancy can be ignored. The SMOS scores indicated that baseline-16kHz has the best speaker similarity performance, followed by DAC-Vocos-16kHz.     

\subsection{Experiment of AR + NAR generation models}
\label{subsec:exp-valle}



\begin{table*}[t]
\centering
\vspace{-2mm}
\caption{Results of AR + NAR models-based speech generation}
\vspace{.3em}
\label{tab:results-VALLE}
\resizebox{\linewidth}{!}{
\begin{tabular}{cccccccccccc}
\toprule
\multirow{2}{*}{ID} & \multirow{2}{*}{Codecs} & \multirow{2}{*}{Token rate} & \multicolumn{3}{c}{Continuation generation}   &  & \multicolumn{5}{c}{Cross-speaker generation}                          \\ \cline{4-6} \cline{8-12} 
                    &                         &                             & SIM-O         & NISQA         & WER (\%)     &  & SIM-O         & NISQA         & WER (\%)    & CMOS      & SMOS      \\ \midrule 
GT                  & GroundTruth             & -                           & 0.67          & 3.87          & 0.96          &  & 0.70           & 3.87          & 0.96          &  0.61         & 4.77          \\ \midrule 
\multicolumn{12}{c}{AR model sampling temperature: 1.0}                                                                                                                                                      \\ \hdashline[1pt/2pt]\hdashline[0pt/1pt]
S1                  & Encodec-official        & 75                          & 0.43          & 3.21          & \textbf{4.73} &  & 0.47          & 3.25          & 3.53          & -1.03          & 4.15          \\
S2                  & Baseline-16kHz          & 50                          & 0.46          & 3.35          & 10.20          &  & 0.48          & 3.32          & 11.25         & -0.73          & \textbf{4.39} \\
S3                  & Baseline-Vocos-16kHz    & 50                          & 0.46          & 3.52          & 10.10         &  & 0.48          & 3.53          & 11.31         & -0.70          & 4.37          \\
S4                  & DAC-16kHz               & 50                          & \textbf{0.48} & \textbf{3.78} & 4.97          &  & \textbf{0.49} & \textbf{3.72} & 3.31          & \textbf{0.01}          &  4.33         \\
S5                  & DAC-Vocos-16kHz         & 50                          & \textbf{0.48} & 3.76          & 4.95          &  & \textbf{0.49} & \textbf{3.72} & \textbf{3.28} & 0.00  &   4.33        \\ \midrule 
\multicolumn{12}{c}{AR model sampling temperature: 0.9}                                                                                                                                                      \\ \hdashline[1pt/2pt]\hdashline[0pt/1pt]
S1a                 & Encodec-official        & 75                          & 0.44          & 3.23          & 3.88          &  & 0.47          & 3.22          & 3.38          & -          & -          \\
S2a                 & Baseline-16kHz          & 50                          & 0.46          & 3.35          & 6.94          &  & 0.48          & 3.31          & 8.52          & -          & -          \\
S3a                 & Baseline-Vocos-16kHz    & 50                          & 0.47          & 3.52          & 6.97          &  & 0.49          & 3.52          & 8.63          & -          & -          \\
S4a                 & DAC-16kHz               & 50                          & \textbf{0.48} & \textbf{3.76} & \textbf{3.08} &  & 0.49          & \textbf{3.75} & 3.03          & -          & -          \\
S5a                 & DAC-Vocos-16kHz         & 50                          & \textbf{0.48} & 3.74          & 3.09          &  & \textbf{0.50} & 3.73          & \textbf{2.91} & -          & -          \\ \midrule   
\multicolumn{12}{c}{AR model sampling temperature: 0.8}                                                                                                                                                      \\ \hdashline[1pt/2pt]\hdashline[0pt/1pt]
S1b                 & Encodec-official        & 75                          & 0.42          & 3.23          & 4.59          &  & 0.45          & 3.16          & 9.80          & -          & -          \\
S2b                 & Baseline-16kHz          & 50                          & 0.46          & 3.35          & 5.39          &  & 0.48          & 3.23          & 8.40          & -          & -          \\
S3b                 & Baseline-Vocos-16kHz    & 50                          & 0.47          & 3.54          & 5.30           &  & 0.48          & 3.46          & 8.55          & -          & -           \\
S4b                 & DAC-16kHz               & 50                          & 0.48          & \textbf{3.74} & \textbf{3.16} &  & \textbf{0.49} & \textbf{3.69} & \textbf{3.82} & -          & -          \\
S5b                 & DAC-Vocos-16kHz         & 50                          & \textbf{0.49} & 3.73          & 3.17          &  & \textbf{0.49} & 3.68          & 3.89          & -           & -           \\   \bottomrule 
\end{tabular}
}
\vspace{-2mm}
\end{table*}

\subsubsection{Dataset}
\label{subsubsec:data-valle}

We trained both the AR and NAR models with 54k hours of Librilight-Large dataset \cite{librilight}. 
For the AR model, data was chunked into samples between 10 and 20 seconds, while for NAR model we used samples of up to 30 seconds. The phoneme data was obtained from the ASR model and the phoneme alignment tool presented in \cite{valle2023} with a frame size of 30ms, whereas the consecutive repetitions of the phonemes were removed. 
For inference, we also used the short-split of LibriSpeech-test-clean \cite{librispeech} data set. 

\subsubsection{Model configuration}
\label{subsubsec:modelCFG-valle}

We adopted the same Transformer architecture as described in the VALL-E work \cite{valle2023} for both AR and NAR models. This architecture consisted of 12 layers, with each layer containing 16 attention heads. The embedding dimension is 1024, and the feed-forward layer dimension is 4096.   

All models were trained for 800k steps with a 6k codec token batch size across 16 GPUs, except the NAR model with Encodec tokens, trained for 560k steps. Using the AdamW optimizer, AR and NAR models underwent optimization with a linear decay learning rate scheduler, 32k warmup steps, and a peak rate of 5e-4.

\subsubsection{Results}
\label{subsubsec:results-valle}

We present the evaluation results in Table~\ref{tab:results-VALLE}. For both continuation and cross-speaker generation, we experimented with three different temperature settings during AR model inference. Notably, different temperatures yielded optimal results for different systems.. For each temperature setting, both DAC-16kHz and DAC-Vocos-16kHz demonstrated equally outstanding performance across all evaluation metrics. An unexpected observation from Table~\ref{tab:results-VALLE} was that the Baseline-16kHz and Baseline-Vocos-16kHz exhibited significantly worse WER performance compared to other codec settings, despite having similar WER for the reconstruction in Table~\ref{tab:results-codec}. 

\begin{figure}[ht]
\vspace{-5mm}
    \centering
    \includegraphics[scale=0.45]{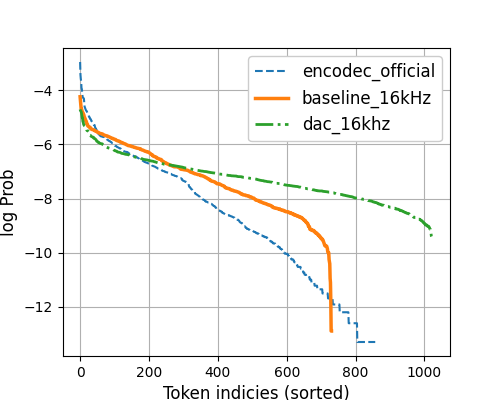}
    \vspace{-2mm}
    \caption{The log-scale distribution of 1st layer codec tokens.} 
\vspace{-2mm}
    \label{fig: codec-dist} 
\end{figure}

To investigate this issue, we analyzed the logarithmic distribution of the first-layer codec token utilization rate using LibriSpeech-test-clean, as depicted in Fig.~\ref{fig: codec-dist}. The Baseline-16kHz curve sharply declined at around 700, indicating the lowest code utilization rate among the three codec systems (Encodec-official and DAC-16kHz used 850 and 1024 codes, respectively). This low code utilization rate likely contributed to the poor WER performance. We infer that this low code utilization is related to the underlying VQ scheme: our baseline model used standard EMA update, while the rest applied techniques to benefit codebook utilization.  
Using a lower temperature during AR model inference significantly improved the WER performance for Baseline-16kHz. The sharper sampling distribution resulting from a lower temperature may mitigate hallucination issues, particularly when the number of used codes is much smaller than the number of classes in the AR model's classification layer. However, we found that the sampling temperature must be within a specific range; an excessively small value could worsen the WER. 

We reported subjective evaluation results when AR model sampling temperature was at 1.0. The same 10 participants from the masked-based parallel speech generation experiments were invited for this test, following the same protocol. The results indicated a positive correlation between CMOS scores and NISQA scores, while SMOS scores fall within a tight range, with Baseline-16kHz achieving the highest score.    

\section{Conclusion}
\label{conclusion}
\vspace{-2mm}

We explored various neural codec systems for the SLM-driven speech generation task. Our investigation included training our own baseline codec systems, as well as variants based on existing high-performing codec systems, including Encodec, Vocos, and DAC. We integrated the codec tokens from these codec systems into two types of SLM-based speech generation systems: masked-based parallel speech generation and AR + NAR models-based text to speech generation systems. Our experiment results revealed that DAC models perform exceptionally well overall for SLM-based speech generation. Vocos served as a competitive vocoder, demonstrating similar performance as the DAC decoder. Interestingly, we observed that the speech reconstruction quality was highly correlated with the naturalness of the generated speech, but this correlation did not hold true for speaker similarity and speech intelligibility.

\newpage

\bibliographystyle{IEEEbib}
\bibliography{strings}

\begin{thebibliography}{10}

\bibitem{GPT2020Brown}
T.~B. Brown, B.~Mann, N.~Ryder, M.~Subbiah, J.~Kaplan, et~al.,
\newblock ``Language models are few-shot learners,''
\newblock in {\em arXiv: 2005.14165}, 2020.

\bibitem{GPT4}
OpenAI,
\newblock ``{GPT}-4 technical report,''
\newblock in {\em arXiv: 2303.08774}, 2024.

\bibitem{audioLM2023}
Z.~Borsos, Raphael Marinier, Damien Vincent, Eugene Kharitonov, Olivier Pietquin, et~al.,
\newblock ``Audio{LM}: a language modeling approach to audio generation,''
\newblock in {\em arXiv: 2209.03143}, 2023.

\bibitem{valle2023}
C.~Wang, S.~Chen, Y.~Wu, Z.~Zhang, L.~Zhou, et~al.,
\newblock ``Neural codec language models are zero-shot text to speech synthesizers,''
\newblock in {\em arXiv: 2301.02111}, 2023.

\bibitem{soundstorm2023}
Z.~Borsos, M.~Sharifi, D.~Vincent, E.~Kharitonov, N.~Zeghidour, and M.~Tagliasacchi,
\newblock ``Soundstorm: Efficient parallel audio generation,''
\newblock in {\em arXiv: 2305.09636}, 2023.

\bibitem{audioPalm2023}
P.~K. Rubenstein, C.~Asawaroengchai, D.~D. Nguyen, A.~Bapna, Z.~Borsos, et~al.,
\newblock ``Audiopalm: A large language model that can speak and listen,''
\newblock in {\em arXiv: 2306.12925}, 2023.

\bibitem{wavenet2016}
A.~van~den Oord, S.~Dieleman, H.~Zen, K.~Simonyan, O.~Vinyals, A.~Graves, N.~Kalchbrenner, A.~Senior, and K.~Kavukcuoglu,
\newblock ``Wavenet: A generative model for raw audio,''
\newblock in {\em arXiv: 1609.03499}, 2016.

\bibitem{vq-wav2vec}
A.~Baevski, S.~Schneider, and M.~Auli,
\newblock ``vq-wav2vec: Self-supervisedlearning of discrete speech representations,''
\newblock in {\em International Conferenceon Learning Representations (ICLR)}, 2020.

\bibitem{hubert}
W.~Hsu, B.~Bolte, Y.~H. Tsai, K.~Lakhotia, R.~Salakhutdinov, and A.~Mohamed,
\newblock ``Hubert: Self-supervised speech representation learningby masked prediction of hidden units,''
\newblock in {\em arXiv:2106.07447}, 2021.

\bibitem{w2v-bert}
Y.~Chung, Y.~Zhang, W.~Han, C.~Chiu, J.~Qin, R.~Pang, and Y.~Wu,
\newblock ``w2v-bert: Combining contrastive learning and masked language modelingfor self-supervised speech pre-training,''
\newblock in {\em IEEE Automatic SpeechRecognition and Understanding Workshop, ASRU}, 2021.

\bibitem{soundstream}
N.~Zeghidour, A.~Luebs, A.~Omran, J.~Skoglund, and M.~Tagliasacchi,
\newblock ``Soundstream: An end-to-end neural audio codec,''
\newblock in {\em arXiv:2107.03312}, 2021.

\bibitem{encodec}
A.~Défossez, J.~Copet, G.~Synnaeve, and Y.~Adi,
\newblock ``High fidelity neural audio compression,''
\newblock in {\em arXiv:2210.13438}, 2022.

\bibitem{dac}
R.~Kumar, P.~Seetharaman, A.~Luebs, I.~Kumar, and K.~Kumar,
\newblock ``High-fidelity audio compression with improved {RVQGAN},''
\newblock in {\em Advances in Neural Information Processing Systems, NeurIPS}, 2023.

\bibitem{gslm_2021}
K.~Lakhotia, E.~Kharitonov, W.-N. Hsu, Y.~Adi, A.~Polyak, et~al.,
\newblock ``On generative spoken language modeling from raw audio,''
\newblock {\em Transactions of the Association forComputational Linguistics}, vol. 9, pp. 1336–1354, 2021.

\bibitem{vq-vae}
A.~van~den Oord, O.~Vinyals, and K.~Kavukcuoglu,
\newblock ``Neural discrete representation learning,''
\newblock in {\em Advances in Neural Information Processing Systems, NeurIPS}, 2017.

\bibitem{vocos}
H.~Siuzdak,
\newblock ``Vocos: Closing the gap between time-domain and fourier-based neural vocoders for high-quality audio synthesis,''
\newblock in {\em International Conference on Learning representations, ICLR}, 2024.

\bibitem{snake-actiFunc}
Z.~Liu, H.~Tilman, and U.~Masahito,
\newblock ``Neural networks fail to learn periodic functions and how to fix it,''
\newblock in {\em Advances in Neural Information Processing Systems, NeurIPS}, 2020.

\bibitem{NaturalSpeech3_2024}
Z.~Ju, Y.~Wang, K.~Shen, X.~Tan, D.~Xin, et~al.,
\newblock ``Naturalspeech 3: Zero-shot speech synthesiswith factorized codec and diffusion models,''
\newblock in {\em arXiv:2403.03100}, 2024.

\bibitem{semantiCodec}
H.~Liu, X.~Xu, Y.~Yuan, M.~Wu, W.~Wang, and M.~D. Plumbley,
\newblock ``Semanti{C}odec: {A}n ultra low bitrate semanticaudio codec for general sound,''
\newblock in {\em arXiv: 2405.00233}, 2024.

\bibitem{promptCodec2024}
Y.~Pan, L.~Ma, and J.~Zhao,
\newblock ``Promptcodec: High-fidelity neural speech codec using disentangled representation learningbased adaptive feature-aware prompt encoders,''
\newblock in {\em arXiv:2404.02702}, 2024.

\bibitem{timeInvCodec2024}
Y.~Ren, T.~Wang, J.~Yi, L.~Xu, J.~Tao, C.~Y. Zhang, and J.~Zhou,
\newblock ``Fewer-token neural speech codec with time-invariant codes,''
\newblock in {\em arXiv:2310.00014}, 2024.

\bibitem{HILCodec2024}
S.~Ahn, B.~J. Woo, M.~H. Han, C.~Moon, and N.~S. Kim,
\newblock ``Hilcodec: High fidelity and lightweight neuralaudio codec,''
\newblock in {\em arXiv:2405.04752}, 2024.

\bibitem{singleCodec2024}
H.~Li, L.~Xue, H.~Guo, X.~Zhu, Y.~Lv, et~al.,
\newblock ``Single-codec: Single-codebook speech codec towards high-performance speech generation,''
\newblock in {\em arXiv:2406.07422}, 2024.

\bibitem{CodecSuperb2024}
Haibin Wu, Ho-Lam Chung, Yi-Cheng Lin, Yuan-Kuei Wu, Xuanjun Chen, et~al.,
\newblock ``Codec-{SUPERB}: An in-depth analysis of sound codec models,''
\newblock in {\em arXiv:2402.13071}, 2024.

\bibitem{reviewALM2024}
H.~Wu, X.~Chen, Y.~Lin, K.~Chang, H.~Chung, et~al.,
\newblock ``Towards audio language modeling - an overview,''
\newblock in {\em arXiv:2402.13236}, 2024.

\bibitem{maskedAudioGen2024}
A.~Ziv, I.~Gat, G.~L. Lan, T.~Remez1, F.~Kreuk1, et~al.,
\newblock ``Masked audio generation using a single non-autoregressive transformer,''
\newblock in {\em International conference on learning represenations (ICLR)}, 2024.

\bibitem{CFG2022}
J.~Ho and T.~Salimans,
\newblock ``Classifier-free diffusion guidance,''
\newblock in {\em arXiv:2207.12598}, 2024.

\bibitem{seanet}
Qitong Wang and Themis Palpanas,
\newblock ``Seanet: A deep learning architecture for data series similarity search,''
\newblock {\em IEEE Transactions on Knowledge and Data Engineering}, vol. 35, no. 12, pp. 12972--12986, 2023.

\bibitem{valle2}
Sanyuan Chen, Shujie Liu, Long Zhou, Yanqing Liu, Xu~Tan, Jinyu Li, Sheng Zhao, Yao Qian, and Furu Wei,
\newblock ``Vall-e 2: Neural codec language models are human parity zero-shot text to speech synthesizers,''
\newblock {\em arXiv preprint arXiv:2406.05370}, 2024.

\bibitem{maskGIT2022}
H.~Chang, H.~Zhang, L.~Jiang, C.~Liu, and B.~Freeman,
\newblock ``Maskgit: Masked generative image transformer,''
\newblock in {\em The IEEE/CVF Computer Vision and Pattern Recognition Conference (CVPR)}, 2022.

\bibitem{librilight}
J.~{Kahn}, M.~{Rivière}, W.~{Zheng}, E.~{Kharitonov}, Q.~{Xu}, P.~E. {Mazaré}, J.~{Karadayi}, V.~{Liptchinsky}, R.~{Collobert}, C.~{Fuegen}, T.~{Likhomanenko}, G.~{Synnaeve}, A.~{Joulin}, A.~{Mohamed}, and E.~{Dupoux},
\newblock ``Libri-light: A benchmark for asr with limited or no supervision,''
\newblock in {\em ICASSP 2020 - 2020 IEEE International Conference on Acoustics, Speech and Signal Processing (ICASSP)}, 2020, pp. 7669--7673,
\newblock \url{https://github.com/facebookresearch/libri-light}.

\bibitem{hifigan}
Jungil Kong, Jaehyeon Kim, and Jaekyoung Bae,
\newblock ``Hifi-gan: generative adversarial networks for efficient and high fidelity speech synthesis,''
\newblock in {\em Proceedings of the 34th International Conference on Neural Information Processing Systems, NeurIPS}, 2020.

\bibitem{librispeech}
Vassil Panayotov, Guoguo Chen, Daniel Povey, and Sanjeev Khudanpur,
\newblock ``Librispeech: An asr corpus based on public domain audio books,''
\newblock in {\em 2015 IEEE International Conference on Acoustics, Speech and Signal Processing (ICASSP)}, 2015, pp. 5206--5210.

\bibitem{pesq}
A.W. Rix, J.G. Beerends, M.P. Hollier, and A.P. Hekstra,
\newblock ``Perceptual evaluation of speech quality (pesq)-a new method for speech quality assessment of telephone networks and codecs,''
\newblock in {\em 2001 IEEE International Conference on Acoustics, Speech, and Signal Processing. Proceedings (Cat. No.01CH37221)}, 2001, vol.~2, pp. 749--752 vol.2.

\bibitem{stoi}
Cees~H. Taal, Richard~C. Hendriks, Richard Heusdens, and Jesper Jensen,
\newblock ``A short-time objective intelligibility measure for time-frequency weighted noisy speech,''
\newblock in {\em 2010 IEEE International Conference on Acoustics, Speech and Signal Processing}, 2010, pp. 4214--4217.

\bibitem{visqol}
Michael Chinen, Felicia S.~C. Lim, Jan Skoglund, Nikita Gureev, Feargus O'Gorman, and Andrew Hines,
\newblock ``Visqol v3: An open source production ready objective speech and audio metric,''
\newblock in {\em 2020 Twelfth International Conference on Quality of Multimedia Experience (QoMEX)}, 2020, pp. 1--6.

\bibitem{mcd}
Robert Kubichek,
\newblock ``Mel-cepstral distance measure for objective speech quality assessment,''
\newblock in {\em Proceedings of IEEE pacific rim conference on communications computers and signal processing}. IEEE, 1993, vol.~1, pp. 125--128.

\bibitem{Chen2021WavLM}
S.~Chen, C.~Wang, Z.~Chen, Y.~Wu, S.~Liu, et~al.,
\newblock ``Wavlm: Large-scale self-supervised pre-training for full stack speech processing,''
\newblock in {\em arXiv:2110.13900}, 2022.

\bibitem{common-voice}
Rosana Ardila, Megan Branson, Kelly Davis, Michael Henretty, Michael Kohler, Josh Meyer, Reuben Morais, Lindsay Saunders, Francis~M. Tyers, and Gregor Weber,
\newblock ``Common voice: A massively-multilingual speech corpus,''
\newblock {\em arXiv preprint arXiv:1912.06670}, 2020.

\bibitem{voicebox2023}
M.~Le, A.~Vyas, B.~Shi, B.~Karrer, L.~Sari, et~al.,
\newblock ``Voicebox: Text-guided multilingual universal speech generation at scale,''
\newblock in {\em arXiv:2306.15687}, 2023.

\bibitem{nisqa2021}
G.~Mittag, B.~Naderi, A.~Chehadi, and S.~Moller,
\newblock ``Nisqa: A deep cnn-self-attention model for multidimensional speech quality prediction with crowdsourced datasets,''
\newblock in {\em arXiv:2104.09494}, 2021.

\end{thebibliography}

\end{document}